\documentclass[12pt,preprint]{aastex}
\usepackage{natbib}
\newcommand{\HI}{\ion{H}{1}} 
\newcommand{\CIV}{\ion{C}{4}}
\newcommand{\OVI}{\ion{O}{6}}

\newcommand{\cm}{{\rm cm}}
\newcommand{\kms}{{\rm km}\,{\rm s}^{-1}}

\begin{document}

\title{The Spatial Distribution of Metals in the Intergalactic Medium}

\author{Matthew M. Pieri\altaffilmark{1,2}, Joop Schaye\altaffilmark{3,4} and
Anthony Aguirre\altaffilmark{5}}

\altaffiltext{1}{Universit\'e Laval, Qu\'ebec QC, G1J 7P4, Canada}
\altaffiltext{2}{Imperial College London, Prince Consort Road, London,
SW7 2AZ, UK} \altaffiltext{3}{Leiden Observatory, PO Box 9513, 2300 RA
Leiden, The Netherlands} \altaffiltext{4}{Institute for Advanced
  Study, Einstein Drive, Princeton, NJ 08540, USA}
\altaffiltext{5}{UC, Santa Cruz, 1156 High
  Street, Santa 
Cruz, CA 95064, USA} 

\begin{abstract}
We investigate the impact of environment on the metallicity of the
diffuse intergalactic medium. We use pixel correlation
techniques to search for weak \CIV\ and \OVI\ absorption in the
spectrum of quasar Q1422+231 in regions of the spectrum close to and
far from galaxies at $z \sim 3$. This is achieved both by using the positions
of observed Lyman break galaxies and by using strong \CIV\ 
absorption as a proxy for the presence of galaxies near the line of
sight.  We
find that the metal line absorption is a strong function of not only the
H\textsc{i} optical depth (and thus gas density) but also proximity to 
highly enriched regions (and so proximity to galaxies). The parameter
``proximity to galaxies'' can account for some, but not all, of the
scatter in the strength of \CIV\ absorption for fixed \HI. Finally, we
find that even if we limit our analysis to the two thirds of the
pixels that are at
least $600~\kms$ from any \CIV\ line that is strong enough to detect 
unambiguously ($\tau_{\mathrm {C\scriptscriptstyle{IV}}} > 0.1$), our statistical analysis
reveals only slightly less \CIV\ for fixed \HI\ than when we analyze the whole spectrum. 
We conclude that while the metallicity is enhanced in
regions close to (Lyman-break) galaxies, the enrichment is likely to
be much more widespread than their immediate surroundings.

\end{abstract}


\maketitle

\section{Introduction}

The mechanical feedback of galaxies (or pre-galactic systems) 
on  the intergalactic medium (IGM) is a central topic in modern
astrophysical cosmology. Outflows from galaxies and pre-galactic systems affect the efficiency of further galaxy formation and evolution by
shock-heating, displacing, and by changing the chemical
composition (and hence the cooling rate) of the intergalactic gas. 
Thus, an understanding of where and to what degree this feedback has
occurred is essential for developing models for galaxy formation and
the evolution of the IGM.

One observable consequence of this type of feedback is the enrichment of the IGM with heavy elements: without this feedback the IGM, which provides the baryon reservoir from which galaxies form, would be entirely free of metals.  Widely distributed metals are indicative of early feedback, but if the enrichment were limited to the environments of
galaxies, late feedback or inefficient mixing would be implied.

The volume filling factor of metal enriched regions is one component
of this spatial distribution and for widespread enrichment a filling
factor as large as 30\% would be expected
(e.g., \citealt{2002ApJ...571...40M}). It has 
been shown that the volume filling factor at $2<z<3$ may be as low as
5\% and still be consistent with observations of \OVI\ in quasar
absorption spectra at $z\sim 2.5$ 
\citep{2004MNRAS.347..985P}.  The degree of metal enrichment of most
of the universe is below the detection limit of current methods and
data. 

Progress has been made, however, by deriving what we can from
the metals that are detected either directly or statistically in
quasar absorption spectra. It has, for example, become clear that 
the IGM metallicity is typically very low ($Z \la 10^{-2} Z_\odot$ at
$z\sim 3$; e.g.,   
\citealt{1995AJ....109.1522C,1996ApJ...465L..95H,1997ApJ...487..482H,2002ApJ...578...43C,2002A&A...396L..11B,2003ApJ...596..768S}), that
there is not much room for evolution of the
column density distribution of \CIV\ from $z\approx 5$ to 2
\citep{2001ApJ...561L.153S} and for evolution of the carbon abundance
from $z\approx 
4$ to 2 \citep{2003ApJ...596..768S}, and that
silicon is overabundant relative to carbon
\citep{2004ApJ...602...38A}. In terms of the {\em distribution} of metals, it has been found that metallicity increases with gas density, but that there is also a large scatter in metallicity at any given density~\citep{2003ApJ...596..768S,2004ApJ...606...92S}.

It seems unlikely that density is the {\em only} environmental factor determining
metallicity: the strong correlations of metal lines with both each
other (e.g., \citealt{2005astro.ph..3001S})
and with Lyman-break Galaxies (LBGs) (e.g., \citealt{astroph/0505122})
suggest that proximity to galaxies may be another important (albeit not independent) factor. This raises two important questions: (1) Are most of the observed metals
confined to the immediate surroundings of galaxies? (2) Can the hidden
parameter ``proximity to galaxies'' account for the large scatter in
the inferred metallicity for fixed density?

One way to answer these questions is to compare the metallicity as a
function of density for two samples of absorbers: those near to
and those far away from galaxies. Assuming, as expected for a photoionized
plasma, that \HI\ Ly$\alpha$ optical depth is a proxy for the gas
density, this measurement would be relatively straightforward if we
could detect and obtain accurate redshifts for most of
the galaxies that are near the line of sight to bright, high-redshift
quasars. In this case we could simply measure metal line absorption as a function of
\HI\ in regions of quasar spectra that are at various distances
from galaxies. However, in practice we cannot easily detect, never
mind obtain accurate redshifts for, the vast majority of galaxies at
high redshift, particularly if they are near to the line of sight
of a much brighter quasar. We are therefore forced to do the next best
thing and either: (1) accept that we are likely missing most
of the galaxies responsible for the observed metal lines and check
whether we can nevertheless see differences between the absorbers either near to
or far from observed galaxies, or (2) use some type of absorption
line as a proxy for galaxies. We have taken both approaches, but found
the latter to be more productive.

We could use strong \HI\ absorption as a proxy for galaxies (this
approach was taken by \citealt{2004A&A...419..811A}). However, this
would complicate the interpretation because \HI\ absorption is
observed to be reduced near an important fraction of starbursting
galaxies \citep{2003ApJ...584...45A,astroph/0505122} and also because we
are already using \HI\ absorption as a proxy for the gas density. We
therefore chose to use strong metal lines instead. These have long been
thought to arise in the immediate vicinity of galaxies (e.g.,
\citealt{1969ApJ...156L..63B,1991A&A...243..344B,2001ApJ...556..158C})
and strong
observational support for this picture comes from 
the recent studies of \cite{2003ApJ...584...45A,astroph/0505122}, who found
that the cross-correlation between \CIV\ lines and LBGs
increases with the strength of the \CIV\ lines and 
becomes comparable to the galaxy autocorrelation function for column
densities $N({\rm CIV}) \ga 10^{12.5}~\cm^{-2}$.

This paper is organized as follows. In the following section we set out the
approach for separation of a quasar absorption spectrum into regions near to
and far from galaxies using LBG positions and strong \CIV\ markers, and describe 
the pixel correlation search for metals in those regions. In
\S\ref{sec:results} 
we present results of the pixel correlation searches, and also
describe an analysis of the scatter in carbon abundance in the
different samples. Finally, we summarize our conclusions in
\S\ref{sec:conclusions}.

\section{Method}
\label{sec:method}
We search for absorption by \CIV\
($\lambda 1548, 1551$\AA) and \OVI\ ($\lambda 1032, 1038$\AA) as
a function of \HI\ ($\lambda 1216, 1026, \ldots, 912$\AA)
in a high-quality spectrum [resolution $6.6~\kms$
  (FWHM), S/N $\ga 100$] of quasar Q1422+231 ($z = 3.62$) taken
with the High Resolution Echelle Spectrograph
\citep{1994SPIE.2198..362V} on the Keck 
telescope, and kindly provided to us by W.~Sargent and M.~Rauch. The
spectrum was reduced as described in \cite{1997AJ....113..136B} and
continuum fitted as described in \cite{2003ApJ...596..768S}. To avoid
proximity effects and confusion with the Ly$\beta$ forest,  
we restrict our analysis to absorption in the redshift range $2.898
\le z \le 3.552$. 

We make use of pixel optical depth statistics 
\citep{1998Natur.394...44C,2002ApJ...576....1A} as implemented in
\cite{2003ApJ...596..768S}. In its simplest form, the pixel technique
involves collating {\mbox Lyman $\alpha$} forest pixels in quasar 
absorption spectra and pairing them with the pixels where absorption
by a chosen metal species is expected. A variety of techniques
are employed to minimise contamination, get around saturation, and to
maximise the information derived in order to arrive at a dependence
between \HI\ Ly$\alpha$ optical depth and apparent optical depths for
\CIV\ and \OVI. We refer to the metal line optical depths as
``apparent'' because they may suffer from some residual contamination 
and noise. These factors raise the
apparent metal line optical depth and simulations indicate that they
adequately explain the value to which this apparent optical depth
asymptotes as $\tau_{\rm HI} \rightarrow 0$ (e.g., 
\citealt{2002ApJ...576....1A, 2004MNRAS.347..985P}).

Statistical error bars are estimated by bootstrap resampling the
observed spectrum after cutting it into chunks of 5~\AA. In the
process we preserve information on whether pixel pairs fall into the
near or far sample. At least 25 pixels and 5 unique chunks must be available 
(i.e., at least 25 pixels must
come from at least 5 different chunks of spectrum)   for 
a given \HI\ bin for the determination of a data point.

To investigate the importance of environmental factors beyond the gas
density, we perform the search on sub-samples of pixels. We split the
samples according to proximity to either observed LBGs or strong \CIV\
absorbers. 

\begin{figure*}[t]
\centering
\includegraphics[scale=.7,angle=90]{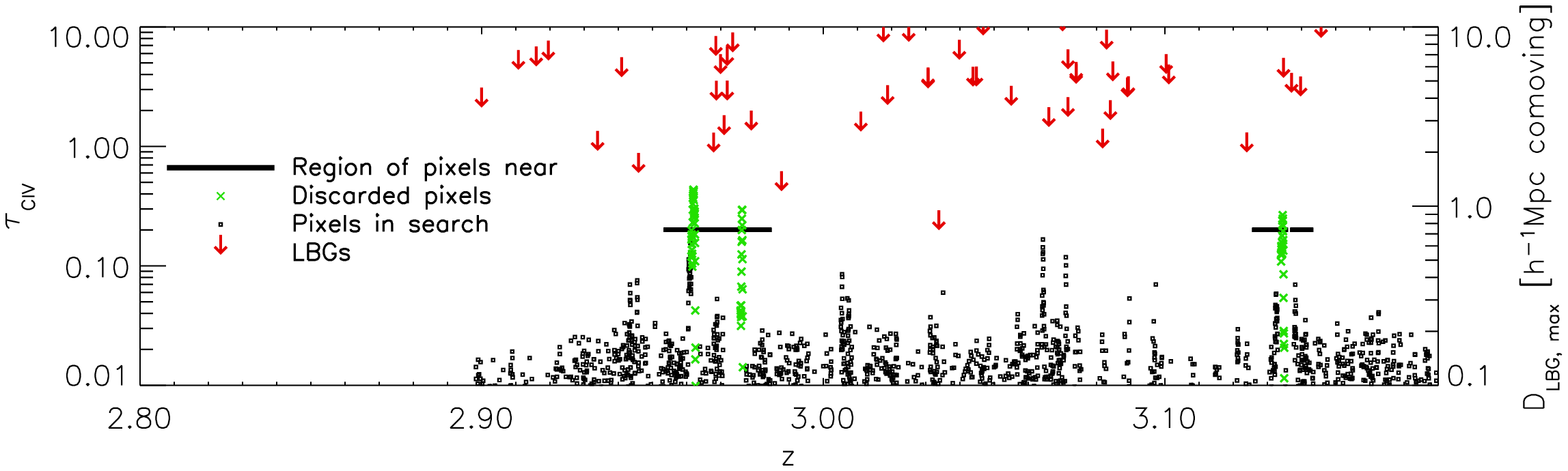}
\includegraphics[scale=.7,angle=90]{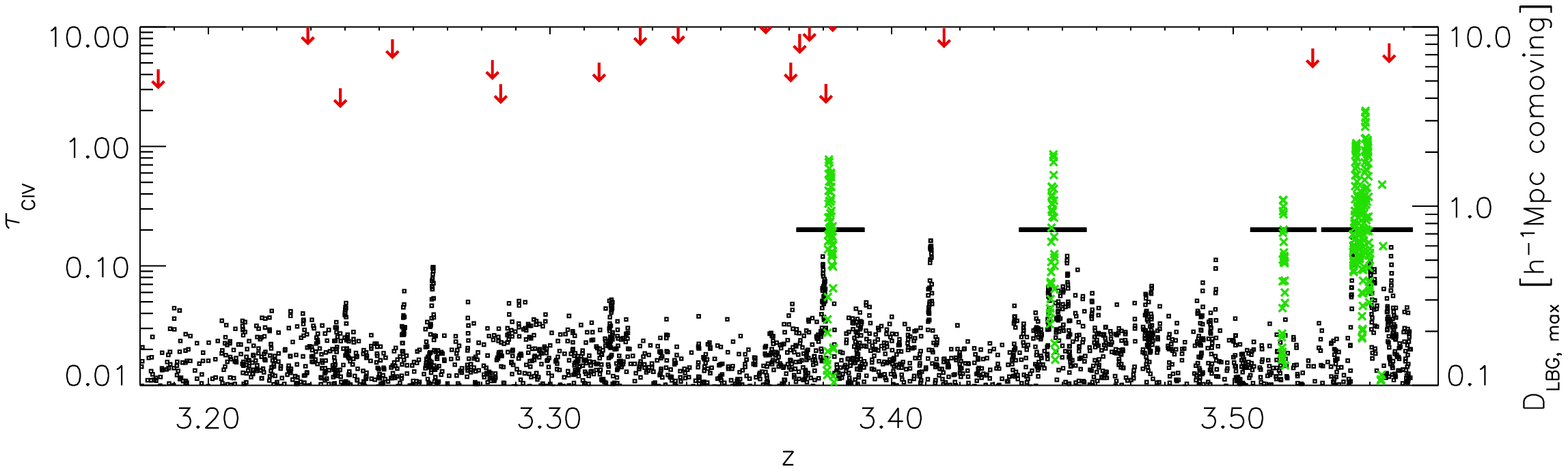}
\caption{\CIV\ pixel optical depth recovered from the spectrum of
  Q1422+231 is shown as a function of redshift (\emph{squares and
  crosses}). The solid lines indicate regions corresponding to the
  near sample for 
$30~\mathrm{kms^{-1}}<|\Delta \upsilon | < 600~ \mathrm{kms^{-1}}$ and
  threshold optical depth $\tau_{\mathrm{C\scriptscriptstyle{IV},
  Thresh}}=0.2$. The solid 
  lines are positioned vertically at the threshold in
$\tau_{\mathrm{C\scriptscriptstyle{IV}}}$. Pixels that are discarded
  because they are within $30~\kms$ of
  marker pixels 
are shown as green crosses.  Positions of LBGs near the line of sight
are indicated with red arrows and their impact parameters can be read off
  from the $y$-axis on the right-hand side.}
\label{civsearchspectrum}
\end{figure*}

The positions and redshifts of LBGs near the line of sight to
Q1422+231 were taken from the online version of Table~18 
of \cite{2003ApJ...592..728S}. We use the procedure of
\cite{2003ApJ...584...45A} to correct the redshifts using $\mathrm
     {Ly\alpha}$ 
emission lines or interstellar absorption lines or both. This provides
an rms scatter in the 
redshift of around $200~\kms$. The locations of the
nearest LBGs are indicated by arrows in
Figure~\ref{civsearchspectrum}, which shows the apparent \CIV\ optical
depth as a function of redshift. The impact parameters of the galaxies
[which were computed assuming $(\Omega_m,\Omega_\Lambda)=(0.3,0.7)$]
can be read off the right $y$-axis. The relative lack of galaxies with $z >
3.4$ reflects the sensitivity function of the selection criteria used
by \cite{2003ApJ...592..728S}. We therefore limit our analysis of
absorption relative to the distance to LBGs to the redshift range
$2.898 < z < 3.25$, over which the LBG completeness is thought to be
more or less uniform \citep{2003ApJ...584...45A}.

Pixels near to and far from LBGs are selected as follows. 
First, LBGs with impact parameter 
$b < D_{\mathrm{LBG, max}}$ relative to the line of sight are each used to
define a marker pixel with redshift $z = z_{\rm LBG}$. Second, 
the full sample of pixels
is separated into those with $| \Delta v |  \equiv |z - z_{\rm LBG}| c
/ (1+z_{\rm LBG}) < 600 ~\kms$ and those with $|\Delta v| >
600~\kms$. Our choice 
for the maximum velocity difference is motivated by the observation
that interstellar absorption lines in the spectra of high-redshift
LBGs are typically blueshifted by up to $600~\kms$ (e.g.,
\citealt{2001ApJ...554..981P,2003ApJ...584...45A}). Such blueshifts
are widely interpreted as a signature of large-scale outflows.

As an alternative to using the LBGs, we also use strong
\CIV\ absorption as a tracer of galaxies, motivated by the findings of
\cite{2003ApJ...584...45A,astroph/0505122}, who compared the
absorption seen in a number of quasar spectra with 
the redshifts of LBGs at small angular separations.
Adelberger et al.\
(\citeyear{astroph/0505122}) conclude that where LBGs are
closer than $\mathrm{\sim 0.25~h^{-1}}$ Mpc
comoving, the average \CIV\ column density is $\sim
10^{14}~\cm^{-2}$. They also find that the galaxy-\CIV\ cross
correlation length increases with $N_{\rm CIV}$, becoming comparable
to the LBG autocorrelation for $N_{\rm CIV} \ga 10^{12.5}~\cm^{-2}$. 
 We use the fact that a \CIV\ Voigt profile line with line width $b$
has a central optical depth of 
\begin{equation}
\tau_{c,{\rm CIV}} = 0.44 \left ({N_{\rm CIV} \over
  10^{13}\,\cm^{-2}}\right ) \left ({ 10~\kms \over b}\right),
\end{equation}
where $b \approx 10~\kms$ is typical for \CIV\ lines (e.g., \citealt{1996ApJ...467L...5R}).
Hence absorbers with $\tau_{\mathrm {C\scriptscriptstyle{IV}}} > 0.1$ should cluster to LBGs as
LBGs cluster to each other, and absorbers with $\tau_{\mathrm {C\scriptscriptstyle{IV}}} \gg 1$
are expected to reside within $10^2$~kpc proper of a LBG.

As with the LBGs, we split the pixels into subsets according to their
velocity relative to the nearest marker
pixel. Marker pixels are in this case defined as those pixels with
$\tau_{\mathrm {C\scriptscriptstyle{IV}}} > \tau_{\mathrm{C{\scriptscriptstyle IV}, Thresh}}$. All
pixels within $\pm 30~\kms$ of a marker pixel are discarded from the
search in order to exclude the possibility that any differences between the two 
samples are trivial consequences of the search method. This would be the 
case if most of the \CIV\ in the near sample of pixels arose from the
absorption lines that contain the marker pixels. In \S\ref{subsec:strongciv}
we will use simulations to show that excluding the nearest $\pm 30~\kms$ is
sufficient for this purpose. Figure~\ref{civsearchspectrum} shows the 
$\tau_{\mathrm {C\scriptscriptstyle{IV}}}$ of those pixels near, far and 
discarded where $\tau_{\mathrm{C{\scriptscriptstyle IV}, Thresh}}=0.2$

\begin{figure}[t]
\centering
\mbox{
\includegraphics[scale=0.36,angle=90]{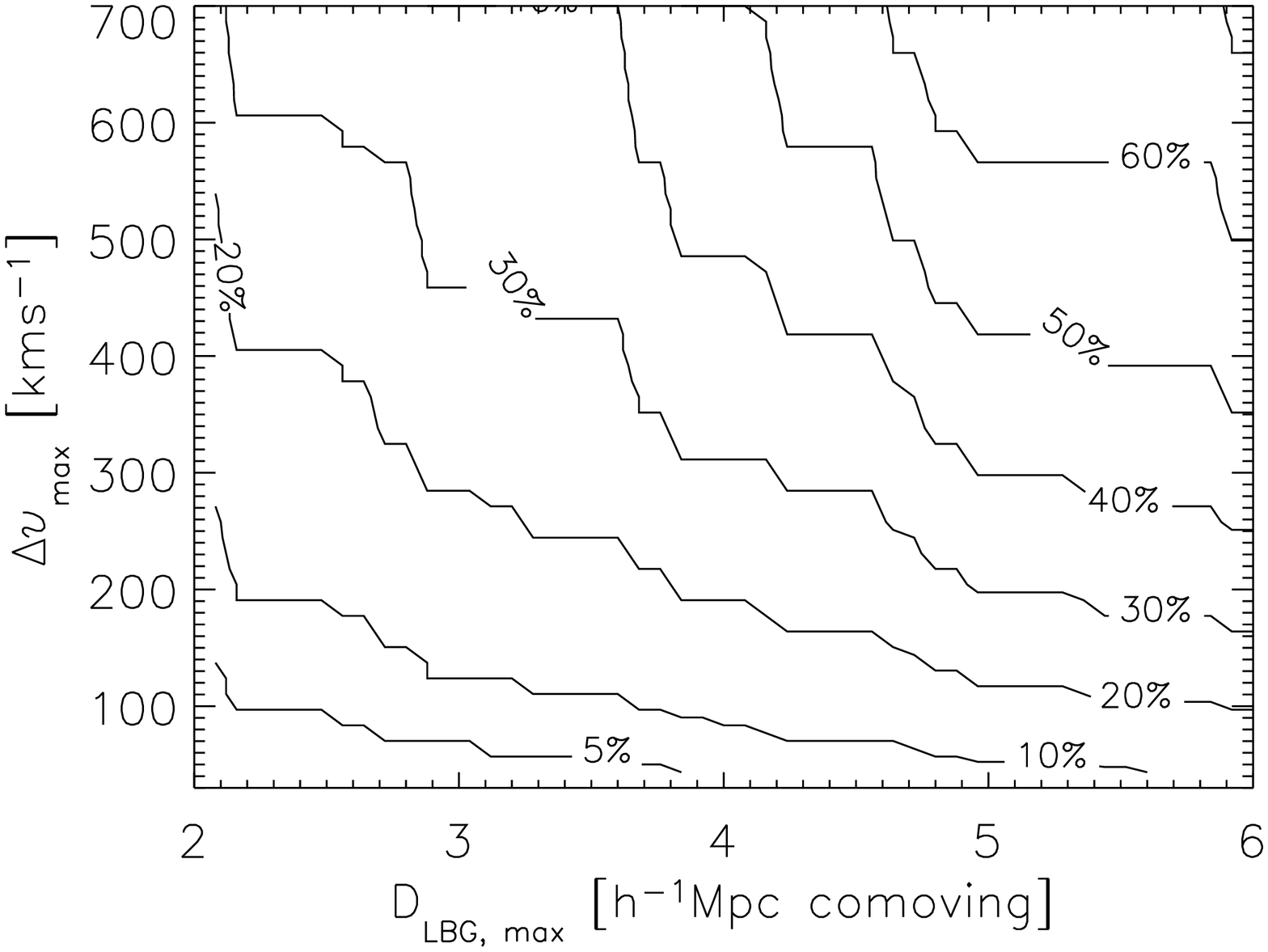}
\includegraphics[scale=0.36,angle=90]{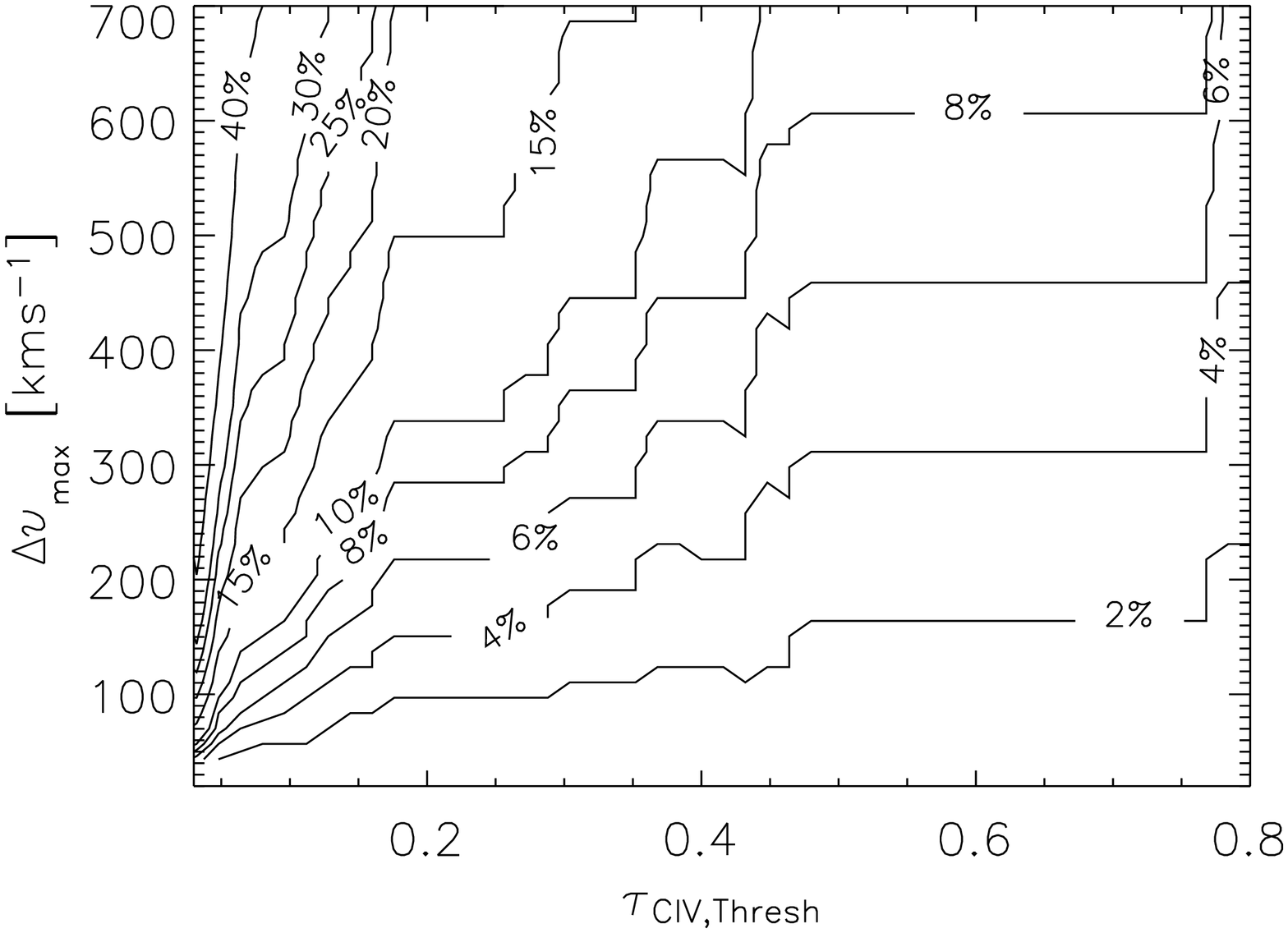}}
\caption{Contour plots showing the fractions of pixel pairs in the
  spectrum of 
Q1422+231 that are within a velocity $|\Delta \upsilon |$ of marker
pixels. \emph{Left:} Marker pixels are defined as those pixels that
have the same redshift as a LBG with impact
parameter $b < D_{\mathrm{LBG, max}}$. Only pixels in the redshift
range $2.898 < z < 3.25$ are considered.
\emph{Right:} Marker pixels
are defined as pixels with recovered optical \CIV\ optical depth
$\tau_{\mathrm{C\scriptscriptstyle{IV}}} > 
\tau_{\mathrm{C{\scriptscriptstyle IV}, Thresh}}$. Only pixels in the
redshift range $2.898 < z < 3.552$ are considered.}
\label{fillingfactors}
\end{figure}

Figure~\ref{fillingfactors} shows the fractions of
pixels that belong to the near
sample when LBGs (\emph{left-hand panel}) or strong \CIV\
(\emph{right-hand panel}) are 
used as markers. A lack of variation in
the fractions of pixels with changing
$D_{\mathrm{LBG,
max}}$ or $\tau_{\mathrm{C{\scriptscriptstyle IV}, Thresh}}$ reflects
a lack of markers in that range. For example there 
are no markers in the range $0.5
<\tau_{\mathrm{C{\scriptscriptstyle IV}, Thresh}}<0.75$ and
$2.2<D_{\mathrm{LBG, max}}<2.5 \mathrm{h^{-1}Mpc}$.

\section{Results}
\label{sec:results}

\subsection{Near to and far from LBGs}

\begin{figure*}[t]
\centering \includegraphics[scale=0.8,
angle=90]{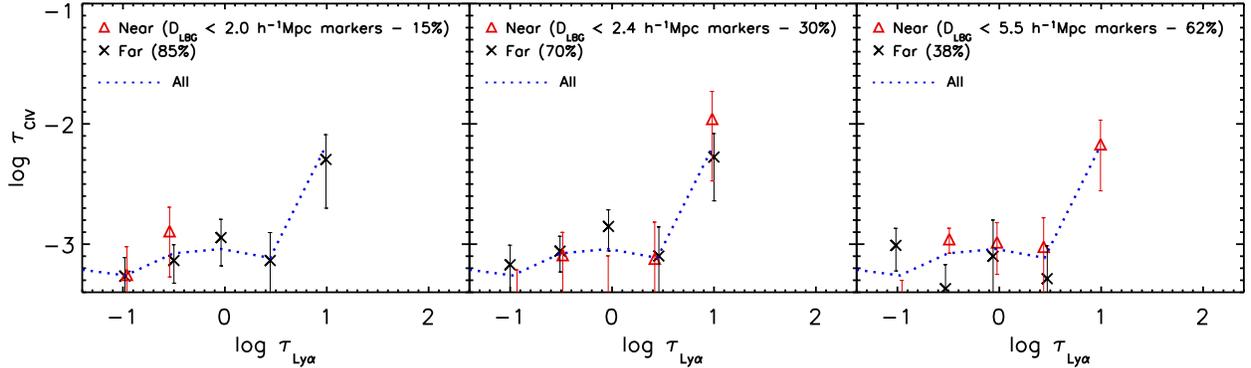}
\caption{$\tau_{\mathrm{Ly\alpha}}$ - median apparent
$\tau_{\mathrm{C\scriptscriptstyle{IV}}}$ relation for Q1422+231
derived using the two sample pixel correlation search for
C\textsc{iv}. Redshifts of LBGs close to the line of sight (impact
parameter $b < D_{\rm LBG}$) are
used as markers for the separation of pixels into samples near to and far
from observed galaxies. The
maximum, comoving impact parameter used is 2.0, 2.4, and
$5.5~h^{-1}\,{\rm Mpc}$ for the left, middle, and right panels
respectively. The near sample (\emph{triangles}) contains all pixels 
with $|\Delta v| < 600~\kms$ from a marker redshift, while the far sample
(\emph{crosses}) is
defined by $|\Delta v| > 600~\kms$. The result 
for the full sample (\emph{dotted lines}) of pixels is shown for comparison.} \label{searchlbg}
\end{figure*}

Figure~\ref{searchlbg} shows the results of the two sample search for
\CIV~\footnote{We do not show the search for weak O{\scriptsize VI} absorption near to and
far from LBGs since the signal-to-noise is low and contamination extensive when 
 searching for O{\scriptsize VI} at redshifts $z<3.25$ in the spectrum of Q1422+231.}
when
the redshifts of nearby LBGs are used as markers, and 
we divide our sample into the ``near"
pixels that are within $600~\kms$ of a marker redshift and the ``far" pixels that
are not. For each panel we take a different maximum LBG impact
parameter $D_{\rm LBG, max}$. 

In the left panel we use the lowest
impact parameter that provides significant results for the near
sample, $D_{\rm LBG, max} = 2 ~h^{-1}\,{\rm Mpc}$ comoving (for smaller
impact parameters the number of pixels and spectral regions in the
near sample becomes too small to satisfy our statistical criteria).
In the middle panel we use $D_{\rm LBG, max} = 2.4~
h^{-1}\,{\rm Mpc}$. This is the scale for which
\cite{2003ApJ...584...45A} found that most of the \CIV\ lines
they detect are seen. In our case seven LBGs are within this distance
from the line of sight.  The right panel shows the case where 
$D_{\rm LBG, max} = 5.5~h^{-1}\,{\rm Mpc}$ comoving. Such an impact parameter 
is chosen to represent the largest useful value for distance to LBGs since, at this redshift,
 this corresponds to $|\Delta v| \approx 600~\kms$ along the line of sight
 just due to the Hubble flow. There are no significant
differences between the two samples for any of the choices of $D_{\rm LBG, max}$. 
This is also the case if we choose to disregard the non-uniform completeness
and use the full redshift range up to $z=3.552$~\footnote{Doing the same in 
a search for O{\scriptsize VI} absorption near and far from LBGs indicates that, 
despite a detection of O{\scriptsize VI} in the sample of all pixels for $\tau_{\rm HI} \ga 10$,
 no signal is seen as the near sample does not extend to high $\tau_{\rm HI}$.}.

This null result is somewhat surprising in light of the findings of
\cite{2003ApJ...584...45A}. However, we emphasize that our redshift
path searched 
is very small ($\Delta z = 0.352$) and that cosmic variance is
therefore likely to be an issue. For example, it can be seen from
Fig.~\ref{civsearchspectrum} that our upper redshift cutoff of
$z=3.25$ (above which 
the photometric selection of LBGs becomes highly incomplete)
results in the
exclusion of many of the stronger \CIV\ absorbers in this spectrum, so that
we do not have sufficient data to investigate the regime $\tau_{{\rm
    Ly}\alpha} > 10$. 
It would thus clearly be desirable to
repeat this analysis using the full sample of
\cite{2003ApJ...584...45A,astroph/0505122}. It should also be
noted that our pixel technique is much more sensitive to weak \CIV\
absorption than the method used by \cite{2003ApJ...584...45A} (i.e.,
direct detection by eye) and much of the absorption seen far from LBGs
may be weak.

Perhaps the most natural explanation for our null result is that we
are missing the vast majority of the galaxies from which the observed
carbon originated. If metals were flowing out of galaxies at a
constant velocity $v_{\rm wind}$, then they would reach a distance
$1.0\times 10^2~{\rm kpc} \left ({v_{\rm wind} \over 10^2~\kms} \right
) \left ({t \over 1 {\rm Gyr}}\right )$ in a time $t$. Hence, it is
natural to expect the sources of the observed intergalactic metals at
$z=3$ to be located well within $1~h^{-1}\,{\rm Mpc}$ comoving of the
line of sight. While we find a large number of \CIV\ absorbers, there
is only one galaxy that satisfies this criterion ($z\approx 3.03$, $b
= 33.66~{\rm arcsec} \approx 0.75~h^{-1}\,{\rm Mpc}$ comoving). This
is perhaps not surprising, given the fact that only the bright end of
the luminosity function is observable and that accurate redshifts can
only be determined for a fraction of these bright galaxies. It does,
however, illustrate the need for a tracer of galaxies that cannot be
detected directly.

\subsection{Near to and far from strong \CIV}
\label{subsec:strongciv}
As discussed in \S\ref{sec:method}, we have carried out a two-sample
search for \CIV\ and \OVI\ absorption in regions near to and far from
strong \CIV\ 
absorbers. This approach was motivated by the work of
\cite{2003ApJ...584...45A,astroph/0505122}, who found that strong
\CIV\ is a good tracer of galaxies.

\subsubsection{\CIV}

\begin{figure*}[t]
\centering \includegraphics[scale=0.8, angle=90]{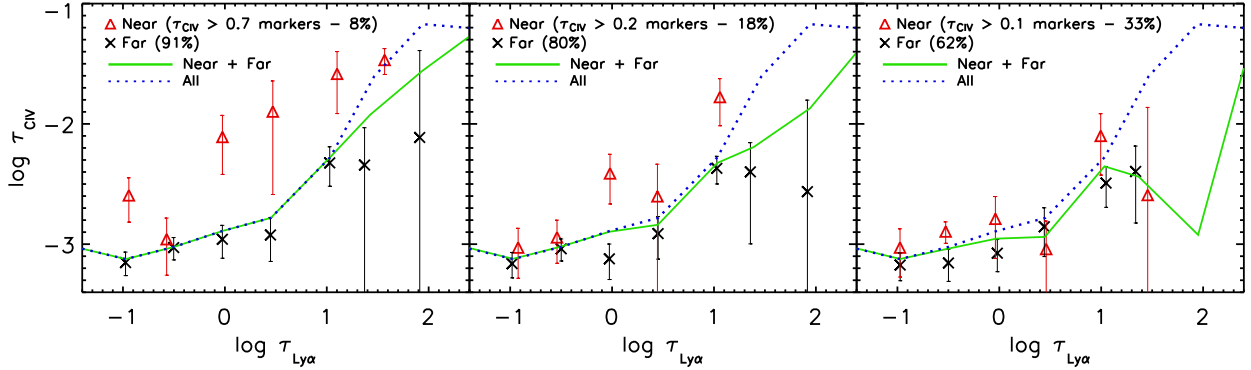}
\caption{$\tau_{\mathrm{Ly\alpha}}$ - median apparent
$\tau_{\mathrm{C\scriptscriptstyle{IV}}}$ relation for Q1422+231
derived using the two sample pixel correlation search for
C\textsc{iv}. Strong C\textsc{iv} absorbers are used as markers 
for the separation of pixels into samples near to and far from relatively
metal-rich regions. 
The threshold optical depth $\tau_{\mathrm{C\scriptscriptstyle{IV},
    Thresh}}$ is 0.7, 0.2 and 0.1 for the left, middle, and right
panels respectively. The near sample (\emph{triangles}) contains all pixels 
with $30~\kms < 
|\Delta v| < 600~\kms$ from a marker redshift, while the far sample
(\emph{crosses}) is
defined by $|\Delta v| > 600~\kms$. The result 
for the combined (\emph{solid lines}) and full (\emph{dotted lines})
samples of pixels are shown for comparison.}
\label{civsearchstrongtau}
\end{figure*}

In Fig.~\ref{civsearchstrongtau} we plot the median, recovered \CIV\
optical depth as a function of $\tau_{\rm HI}$ for pixels near
(\emph{triangles}) 
and far (\emph{crosses}) from strong 
\CIV\ absorbers. For comparison the results for the combined near and
far samples (\emph{solid lines}) and for the full sample of pixels
(\emph{dotted 
lines}) are also shown. Comparison with Fig.~\ref{searchlbg} shows that the near
doubling of the redshift path searched ($\Delta z = 0.654$ vs.\
$\Delta z = 0.352$) significantly improves the statistics,
particularly for high $\tau_{\rm HI}$.

The full sample and the combined near and far sample differ because only the full 
sample contains those pixels that are within $30~\kms$ of a marker
pixel along with the marker pixel itself. As the threshold is lowered
the results from these two samples become increasingly different as more
pixels leave the combined near and far sample. The effect is pronounced
for pixels with $\tau_{\rm HI}>10$ but is small for $\tau_{\rm HI}<10$. Since
the systems of interest in this study are those with $\tau_{\rm HI}<10$
we can confirm that increasing the number of discarded
pixels does not play a significant role.

The left, middle, and right panels use threshold
optical depths,  
$\tau_{\mathrm{C{\scriptscriptstyle IV}, Thresh}}$, of 0.7, 0.2, and
0.1, respectively, which covers the interesting range of values.
Using 
$\tau_{\mathrm{C{\scriptscriptstyle IV}, Thresh}} > 0.7$, would leave only  
the lowest \HI\ bins in the near sample with a sufficient number of
pixels to satisfy our statistical criteria.  The lower limit,
$\tau_{\mathrm{C{\scriptscriptstyle IV}, Thresh}} = 0.1$, corresponds
to the lowest optical depth which we can unambiguously identify as 
being due to \CIV, and also corresponds to the central optical depth
above which \CIV\ absorption lines cluster to LBGs as LBGs cluster to
each other (see \S\ref{sec:method}). In addition, the number of pixels
with $\tau_{{\rm Ly}\alpha}\gg 1$ becomes insufficient for
$\tau_{\mathrm{C{\scriptscriptstyle IV}, Thresh}} < 0.1$ 
because in that case  many pixels
fall within $30~\kms$ of a marker pixel. 

In each panel the near sample consists of all pixels
that are at most $600~\kms$ from the nearest marker pixel, which are defined by
$\tau > \tau_{\mathrm{C{\scriptscriptstyle IV}, Thresh}}$, and at least $30~\kms$ from all
marker pixels. The far sample always consists of the pixels that are
at least $600~\kms$ from all marker pixels. The fractions of pixels
contained in the near and far samples are indicated in the panels, the
fractions do not sum to exactly 100\% because the combined sample
still excludes those pixels that are within $30~\kms$ of a marker
pixel. 

It is clear that, for fixed
$\tau_{\rm HI}$, the \CIV\ absorption is much
stronger for pixels in the
vicinity of $\tau_{\mathrm {C\scriptscriptstyle{IV}}} > 0.7$ and 0.2 absorbers than for pixels
that are at least $600~\kms$ away from such marker pixels. Since
$[{\rm C}/{\rm H}] \propto \log\tau_{\mathrm {C\scriptscriptstyle{IV}}}$ for fixed $\tau_{\rm HI}$
if the gas if photoionized, we see that in these regions carbon may be
overabundant by up to an order of magnitude compared to the global
median. However,
when we lower the threshold optical depth from 
$\tau_{\mathrm {C\scriptscriptstyle{IV}}} = 0.2$ to 0.1, and thus increase the fraction of
pixels in the near sample from 18 to 33 percent, the differences
between the samples become insignificant. These results show that the
\HI\ optical depth, which is thought to be a good indicator of the gas
density, is not the only parameter that correlates with
metal enrichment. Proximity to highly enriched regions and (by
extension) galaxies clearly also plays an important role.

Interestingly, the results
for the far sample are always consistent with those for the combined
samples and disagree with the full sample only for the highest \HI\
bins ($\tau_{\rm HI} \gg 10$), where the full sample becomes dominated
by pixels  
within $30~\kms$ of a $\tau_{\mathrm {C\scriptscriptstyle{IV}}} > 0.1$ marker pixel.
The median
apparent $\tau_{\mathrm {C\scriptscriptstyle{IV}}}$ is always substantially higher for 
$\tau_{\rm HI} \ga 10$ than for $\tau_{\rm HI} \ll 10$, even if we
only consider the 62\% of the pixels that are at least $600~\kms$ from
any pixels that have $\tau_{\mathrm {C\scriptscriptstyle{IV}}} > 0.1$
(crosses in the right-hand panel). Thus, it appears that carbon
enrichment is \emph{not} confined to the immediate surroundings of
highly enriched regions.  

\begin{figure*}[t]
\centering
\includegraphics[scale=.8,angle=90]{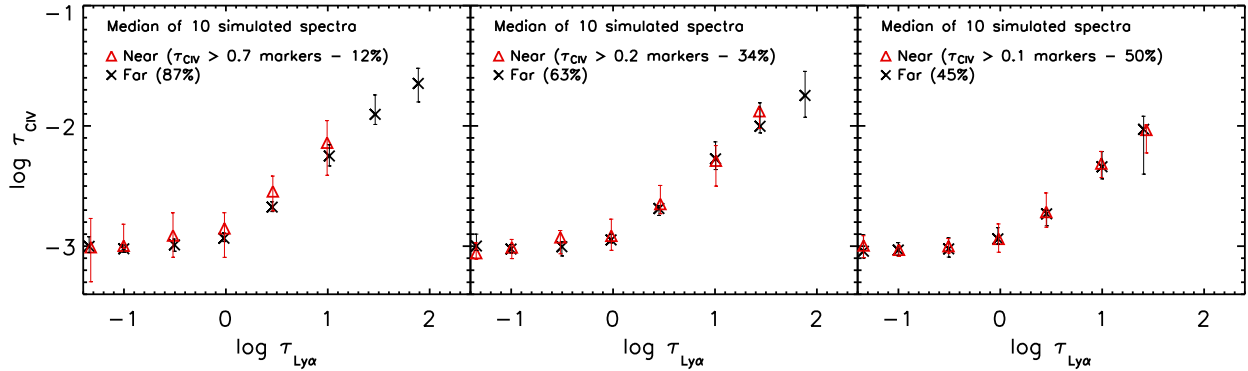}
\caption{$\tau_{\mathrm{Ly\alpha}}$ - median apparent
$\tau_{\mathrm{C\scriptscriptstyle{IV}}}$ relation for 10 simulated
spectra derived using the two sample pixel correlation search for
C\textsc{iv} as used in Fig.~\protect\ref{civsearchstrongtau}. Data
points indicate the medians of the 10 spectra. The synthetic spectra
were taken from a cosmological, hydrodynamical simulation in which the
metals are distributed according to the measurements of
\protect\cite{2003ApJ...596..768S}. The metal distribution is
stochastic and depends only on the gas density and redshift.}  
\label{searchsim} 
\end{figure*}

At this point some scepticism would certainly be justified. Since we
are looking for \CIV\ in regions selected by 
\CIV, aren't our results a trivial consequence of the search
method? Figure~\ref{searchsim} demonstrates that is not the case. The figure
is identical to Fig.~\ref{civsearchstrongtau}, except that
we have now plotted the results for spectra drawn from a
cosmological, hydrodynamical simulation in which the metals were
distributed according to the observations of
\cite{2003ApJ...596..768S}, who measured the distribution of
carbon (including the scatter) as a function of density and redshift,
as inferred from \CIV\ 
absorption in a large set of quasar spectra. The simulation, the
procedures for creating a spectrum that mimics the observations, and
the metal distribution are all described in detail in
\cite{2003ApJ...596..768S}; for present purposes the key point is that
the metallicity assigned to the gas depends on the gas density but
{\em not} explicitly on the gas's proximity to galaxies.  
Clearly, splitting the pixels into
samples near to and far from strong \CIV\ absorbers has no effect for
these synthetic spectra. This demonstrates that the significant
differences between the two observed samples cannot be explained as an
artifact of the search method.

It was shown by \cite{2003ApJ...596..768S} that, for fixed $\tau_{\rm
  HI}$, the $\tau_{\mathrm {C\scriptscriptstyle{IV}}}$ 
distribution in Q1422+231 is well-fit by a lognormal distribution of
width $\sigma(\log \tau_{\mathrm {C\scriptscriptstyle{IV}}}) = \sigma([{\rm C}/{\rm H}]) \simeq
0.8$. 
Now that we have established that proximity to highly enriched gas is a 
second parameter, along with density, that controls the amount of \CIV\
absorption, it is interesting to
ask whether this parameter is responsible for the large amount of
scatter in $\tau_{\mathrm {C\scriptscriptstyle{IV}}}$ for fixed $\tau_{\rm HI}$. To
test this, we have determined $\sigma(\log\tau_{\mathrm {C\scriptscriptstyle{IV}}})$ near to and
  far from strong \CIV\
absorbers. 

Briefly, the method used is as follows (see
  \citealt{2003ApJ...596..768S} for more detail).
For each $\tau_{{\rm Ly}\alpha}$ bin, we
calculate many percentiles $\log\tau_{\mathrm {C\scriptscriptstyle{IV}}}(s)$, where $s$ is the
number of standard deviations in a Gaussian distribution; for example,
$\log\tau_{\mathrm {C\scriptscriptstyle{IV}}}(s=0)$ and $\log\tau_{\mathrm {C\scriptscriptstyle{IV}}}(s=1)$ would,
respectively, correspond to the 50th and 84th percentiles in the
$\log\tau_{\mathrm {C\scriptscriptstyle{IV}}}$ distribution. Errors on these points are
calculated using the same bootstrap procedure as described in
\S\ref{sec:method}. From each of these, we subtract $\log\tau_{\rm
CIV}(s)$ computed from all pixels (in the full undivided sample) with
$\tau_{{\rm Ly}\alpha} < -0.5$; in this way, we subtract off the effect
  of noise and 
contamination that is uncorrelated with $\tau_{{\rm Ly}\alpha}$. Next,
we fit a line to the function $\log\tau_{\mathrm {C\scriptscriptstyle{IV}}}(s)$, which
corresponds to a lognormal fit to the $\tau_{\mathrm {C\scriptscriptstyle{IV}}}$ probability distribution; the slope of the
linear fit to $\log\tau_{\mathrm {C\scriptscriptstyle{IV}}}(s)$ is the width $\sigma(\log\tau_{\rm
CIV})$ of the lognormal distribution, in dex. To obtain errors we
  repeat the procedure using bootstrap-resampled 
realizations of the spectrum.

\begin{figure}[t]
        \plotone{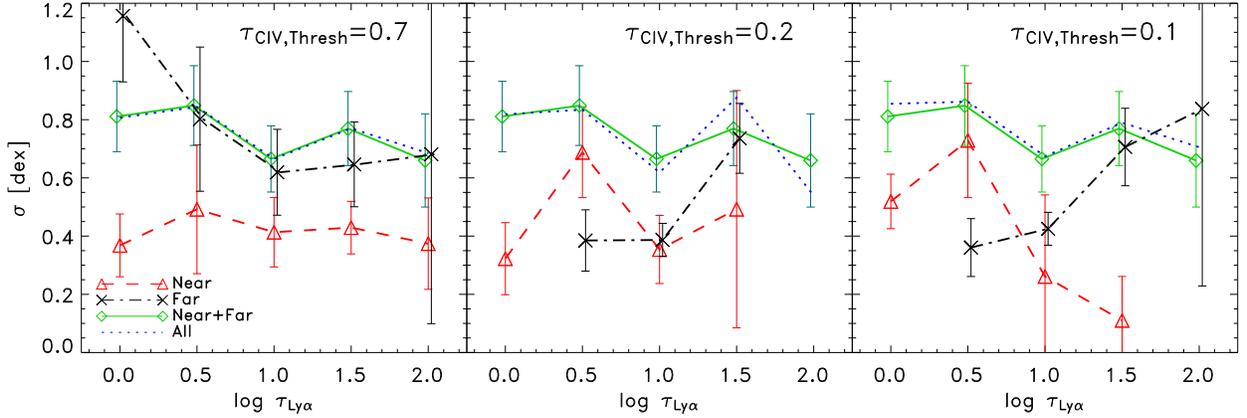}
        \figcaption{Scatter in the \CIV\ abundance for various samples of pixels. Each point corresponds to the scatter $\sigma$ in
        $\log\tau_{\mathrm {C\scriptscriptstyle{IV}}}$, in dex, for a bin of $\log\tau_{{\rm
        Ly}\alpha}$. The samples are identical to those in
        Fig.~\protect\ref{civsearchstrongtau}. 
        \label{fig-scatfig}}
\end{figure}

The results are shown in Figure~\ref{fig-scatfig} for the threshold
\CIV\ optical depths 0.7 (\emph{left}), 0.2 (\emph{middle}), and 0.1
(\emph{right}). In
all cases there is significantly less scatter in the near sample
than in the full sample. In addition, for $\tau_{\mathrm{C{\scriptscriptstyle IV}, Thresh}} =
0.2$ and 0.1, there is some evidence that
the scatter in the far sample is {\em also} smaller. These results suggest
that variations in the distance to the nearest galaxy is responsible
for a substantial fraction of the scatter in the metallicity of gas at a
fixed density.

\subsubsection{\OVI}

\begin{figure*}[t]
\centering \includegraphics[scale=0.8,
angle=90]{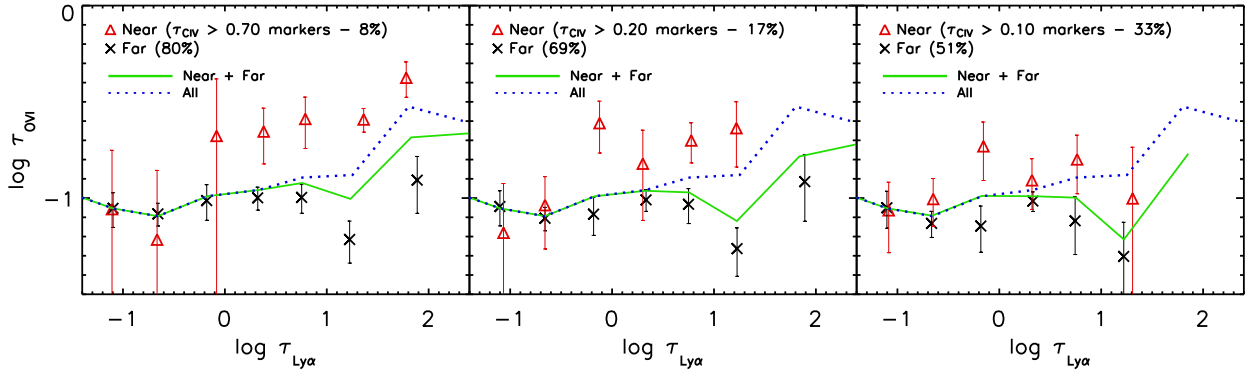}
\caption{As Fig.~\protect\ref{civsearchstrongtau}, except that we
  search for O\textsc{vi} instead of C\textsc{iv}. The pixels are,
  however, still 
  divided into samples near to and far from strong C\textsc{iv}}.
\label{ovisearchstrongtau}
\end{figure*}

The strong C\textsc{iv} markers can of course also be used in
searches for other ions. Figure~\ref{ovisearchstrongtau}
shows the results for \OVI. The redshift range searched, the threshold
\CIV\ optical depths, and velocity range are all the same as in the
\CIV\ search. The small differences in the fractions of
pixels belonging to the various samples between
Figs~\ref{civsearchstrongtau} and \ref{ovisearchstrongtau} are
due to differences between the numbers of \OVI\ and \CIV\ pixels
discarded due to noise spikes, sky lines, and contamination.

As was the case for \CIV, there are significant differences between
the near and far samples for $\tau_{\rm CIV,Thresh} = 0.7$ and
0.2, although the differences are much smaller than was the case for \CIV\
($\sim 0.5$ dex rather than 1 dex). In contrast to \CIV, we find no
evidence for \OVI\ absorption in 
any of the far samples, except perhaps tentatively in the very
highest \HI\ bin ($\tau_{\rm HI} \sim 10^2$). These results indicate
that either oxygen enrichment is less widespread than carbon, or that
away from the highly enriched regions oxygen is too weak to
detect. The latter scenario appears more 
plausible and would also explain why the samples differ less in $\tau_{\rm
  OVI}$ than in $\tau_{\mathrm {C\scriptscriptstyle{IV}}}$ (even if the difference in
the true, median $\tau_{\rm OVI}$ between the near and far samples
were as large as a factor ten, we would not be able to tell if the
median for the near sample is only $\sim0.5$ dex above the detection
limit).  

It would be interesting to
repeat this analysis at $z\sim 2.5$, where the pixel search has been
shown to be able to produce a strong \OVI\ signal
\citep{2000ApJ...541L...1S}. \cite{2004A&A...419..811A}
have in fact performed a two-sample pixel search for \OVI\ at this
redshift, but using strong \HI\ absorbers as markers rather than
strong \CIV. They found that
for $\log \tau_{\rm HI} \approx -0.5$ the corresponding \OVI\ is
only detectable in the near sample, which in their case consisted of
all pixels that are less than $500~\kms$ from a pixel with $\tau_{\rm
  HI} > 4$ (they do not give the fraction of pixels represented by
this subset, but we note that for Q1422+231 their choice of parameters
would put 88\% of the pixels in the near sample).

\section{Conclusions}
\label{sec:conclusions}

We have searched for absorption by \CIV\ and \OVI\ absorption at $2.898
\le z \le 3.552$ in a high-quality spectrum of quasar Q1422+231.
We can summarize our findings regarding the spatial distribution of metal
enrichment in the intergalactic medium as follows:

\begin{itemize}
\item No correlations between metal line absorption and
the positions of Lyman-break galaxies are found. This may not be
inconsistent with findings by others based on substantially
larger samples which focused on stronger \CIV\ absorbers. This part of
our analysis
was restricted to the redshift range
$2.898<z<3.25$ - a comparatively narrow range required by the need for
consistent completeness in the detection of LBGs.

\item We find a strong correlation between metal line absorption and
the location of strong \CIV\ absorbers, which
\cite{2003ApJ...584...45A,astroph/0505122} found to be good tracers
of galaxies. Two-sample searches for \CIV\ and \OVI\ near and far from
strong \CIV\ absorbers demonstrate
that the enrichment depends not only on the density, but also on the
proximity to regions that are highly enriched (and so proximity to 
galaxies).

\item The variations in proximity to highly enriched regions
can account for a substantial part of the scatter in the abundance of
carbon for a fixed \HI\ strength.

\item In searching for weak \CIV\ it is clear
that the enrichment is much more widespread than the regions
surrounding strong metal-line absorption, but for \OVI\ the
detection limit  (which is set by contamination from the \HI\ Lyman
series) is too high
to probe the enrichment in all but the most metal rich regions.

\item Metal enrichment is seen both far from currently detectable
   galaxies and far from strong \CIV\ absorbers.

\end{itemize}

\bigskip 
\begin{acknowledgments}

We would like to gratefully acknowledge Wal Sargent and Michael Rauch
for providing access to the observed spectrum Q1422+231, Tom Theuns
for the simulation used for Figure~\ref{searchsim} and Martin Haehnelt for 
helpful discussions. M.P. thanks 
the Canada Research Chair program and NSERC for support. This work was
partially supported by a EU Marie Curie Excellence Grant
(MEXT-CT-2004-014112), the 
W.M.~Keck Foundation, the National Science Foundation (PHY-0070928),
and by  the European Community Research and Training
Network of ``The Physics of the Intergalactic Medium''. M.P. would also like to 
thank Max-Planck-Institut f\"ur Astrophysik for their hospitality.

\end{acknowledgments}

\end{document}